\documentclass[pra,twocolumn,nofootinbib,preprintnumbers,superscriptaddress]{revtex4-2}

\usepackage{natbib}
\usepackage{amssymb,amsbsy,amsmath,amsfonts}
\usepackage{mathrsfs}
\usepackage{graphicx}% Include figure files
\usepackage{epsf,epsfig,latexsym,amsthm,fancyhdr,rotating}
\usepackage{graphics,psfrag,longtable}
\usepackage{slashed}
\usepackage{esint}
\usepackage{nicefrac}
\usepackage{braket}
\usepackage{mathtools}
\usepackage{ulem}

\newcommand{\be}{\begin{equation}}
\newcommand{\ee}{\end{equation}}

\usepackage{color}
\definecolor{purple}{rgb}{.36,.12,.60}
\definecolor{orange}{rgb}{.9,.3,.0}
\definecolor{green}{rgb}{0,.6,0}
\definecolor{RED}{rgb}{.9,.3,.1}

\usepackage[
    colorlinks=true,
    citecolor=blue,
    linkcolor=blue,
    urlcolor=blue,
    breaklinks=true
]{hyperref}

\usepackage{xcolor}

\newcommand{\bluecite}[1]{\textcolor{blue}{\cite{#1}}}

\renewcommand{\doi}[1]{%
  \href{https://doi.org/#1}{doi:#1}%
}

\newcommand{\arxiv}[1]{%
  \href{https://arxiv.org/abs/#1}{arXiv:#1}%
}

\usepackage[percent]{overpic}
\usepackage{hypcap}
\usepackage{placeins}
\usepackage{float}

\usepackage{placeins}

\begin{document}

\title{Diffractive Two-Photon Exchange and Beam Normal-Spin Asymmetries \\ for Elastic Electron Scattering on Nuclei}

%\title{Optical and Fermion Equivalents in Polarization Textures of Vortex Waves}

%LGian Beams of Relativistic FermionsLGian Beams of Relativistic Fermions

\author{Volodymyr Tereshchuk}
\author{Andrei Afanasev}
\affiliation{Department of Physics,
The George Washington University, Washington, DC 20052, USA}

\begin{abstract}
We developed a theoretical approach to elastic electron scattering on nuclei that includes a two-photon-exchange mechanism responsible for parity-conserving single-spin beam asymmetries. The two-photon-exchange amplitude at small scattering angles is treated in a diffractive framework similar to that of pion-nucleus elastic scattering applied to the nuclei $^{12}_{6}$C, $^{40}_{20}$Ca and $^{208}_{82}$Pb. The predicted kinematic features of the beam polarization asymmetries for different nuclei may reconcile Jefferson Lab's experimental results at finite scattering angles with a forward limit given by an optical theorem, potentially resolving the so-called ``PREX Puzzle" for $^{208}_{82}$Pb. 
\end{abstract}
\date{\today
}
\maketitle

%%%%%%%%%%%%%%%%%%%%%%%%%%%%%%%%%%

%%%%%%%%%%%%%%%%%%%%%%%%%%%%%%%%%%

%%%%%%%%%%%%%%%%%%%%%%%%%%%%%%%%%%

\section{Introduction}

Over the past several years, two-photon exchange (TPE) effects have been a topic of considerable study \bluecite{carlson2007,arrington2011,afanasev2017,borisyuk2009,metz2012,koshchii2019,schmidt2023}. The existing discrepancy between the extracted ratios of the electric-to-magnetic Sachs form factors, $G_E/G_M$ \bluecite{kelly2004}, obtained using Rosenbluth separation \bluecite{rosenbluth1950,andivahis1994} and polarization-transfer techniques \bluecite{jones2000,gayou2002}, is one of the reasons why the proper treatment of TPE effects is of great interest. The latter technique is less sensitive to higher-order electromagnetic effects, making it reasonable to assume that the TPE corrections to the Born cross section could account for the observed experimental discrepancy, see \bluecite{afanasev2017} for review. This subject of research has generated growing interest in both theoretical \bluecite{afanasev2004prd,afanasev2004plb,pasquini2004,gorchtein2006,gorchtein2008,carlson2017,ahmed2023} and experimental studies \bluecite{wells2001,maas2005,armstrong2007} of beam-normal single-spin asymmetries (BNSSAs), which provide direct access to the imaginary part of the TPE amplitude.

Another reason that emphasizes the importance of studying BNSSAs is the role they play in parity-violating (PV) experiments. PV asymmetries serve as a significant testing ground for the electroweak theory \bluecite{qweak2018}, and, importantly, are used to determine the neutron skin thickness \bluecite{horowitz2012,abrahamyan2012skin}, which in turn helps to constrain the nuclear equation of state and facilitate the theoretical description of neutron-rich matter. PV asymmetries are measured by scattering a longitudinally polarized beam from an unpolarized target. Typically, this observable is of the order of $0.5$ ppm, whereas typical beam-normal SSAs are $5-20$ ppm, which means that the presence of even a minor transverse polarization component may induce significant systematic errors in the measurements of PV asymmetries. For this reason, accurate predictions of BNSSA values for various nuclear targets remains one of the major goals of modern theoretical studies in precision electron scattering.

The existing theoretical models rely on the optical theorem to determine the results in the forward limit and then introduce a phenomenological $t$-dependence of the Compton amplitude to calculate the asymmetries at nonzero scattering angles \bluecite{gorchtein2008,koshchii2021}. They are effective for the description of BNSSA in the case of low-Z nuclei. However, recent measurements of beam-normal SSA for ${}^{208}_{82}\mathrm{Pb}$ performed by the HAPPEX and PREX Collaborations and later by the CREX and PREX Collaborations at Jefferson Lab indicate significant discrepancies \bluecite{abrahamyan2012bnssa,adhikari2022}, termed the ``PREX puzzle," between the experimental results and all existing theoretical predictions throughout a broad spectrum of the squared momentum transfer, $Q^2$. Until recently, it was presumed that the discrepancy could be attributed to Coulomb distortions overlooked in current methodologies \bluecite{cooper2005}. However, recent studies indicate that Coulomb distortions cannot fully account for the observed mismatch \bluecite{koshchii2021,jakubassa2023,jakubassa2024}. Despite significant efforts, the PREX puzzle remains largely unexplained, which reinforces the need for further research. Recent measurements at MAMI \bluecite{esser2025pb} produced the asymmetry on ${}^{208}_{82}\mathrm{Pb}$ consistent in magnitude and sign with measurements on light nuclei which may indicate significant dependence of the asymmetry on the kinematics.

In this study, we consider a diffractive framework for modeling the dependence of the Compton amplitude on the squared momentum transfer \bluecite{chen1993,kahrimanis1997}. We start with the calculation of the imaginary parts of the amplitudes for the elastic scattering of pions, $\pi^{\pm}$, from a nuclear target \bluecite{chen1993,kahrimanis1997}. Next, we use the additive quark model (AQM) \bluecite{lipkin1966} to determine the imaginary part of the $\rho$-meson scattering amplitude. As a final step, we apply the vector-meson-dominance (VMD) model \bluecite{bauer1978,deppman2006} to infer the behavior of the Compton amplitude for nonzero values of $t$. Our results suggest that the notable differences between the measured BNSSA values for ${}^{208}_{82}\mathrm{Pb}$ and those for other spin-zero nuclei, namely the positive sign of the asymmetry and its near-zero values, can potentially be captured within this framework. Small positive values of the observable may be explained by the combination of diffractive minima in the Compton amplitude occurring near the momentum transfer values probed in the experiments and the diffractive minimum of the lead charge form factor \bluecite{duda2007}, which together cause the oscillatory behavior of the asymmetry over the $t$ region covered by the experiments.

%%%%%%%%%%%%%%%%%%%%%%%%%%%%%%%%%%

%%%%%%%%%%%%%%%%%%%%%%%%%%%%%%%%%%

\begin{figure}[t]
    \capstart
    \centering

    \begin{minipage}[t]{0.42\columnwidth}
        \centering
        \scalebox{1}[0.99]{%
        \includegraphics[width=\linewidth]{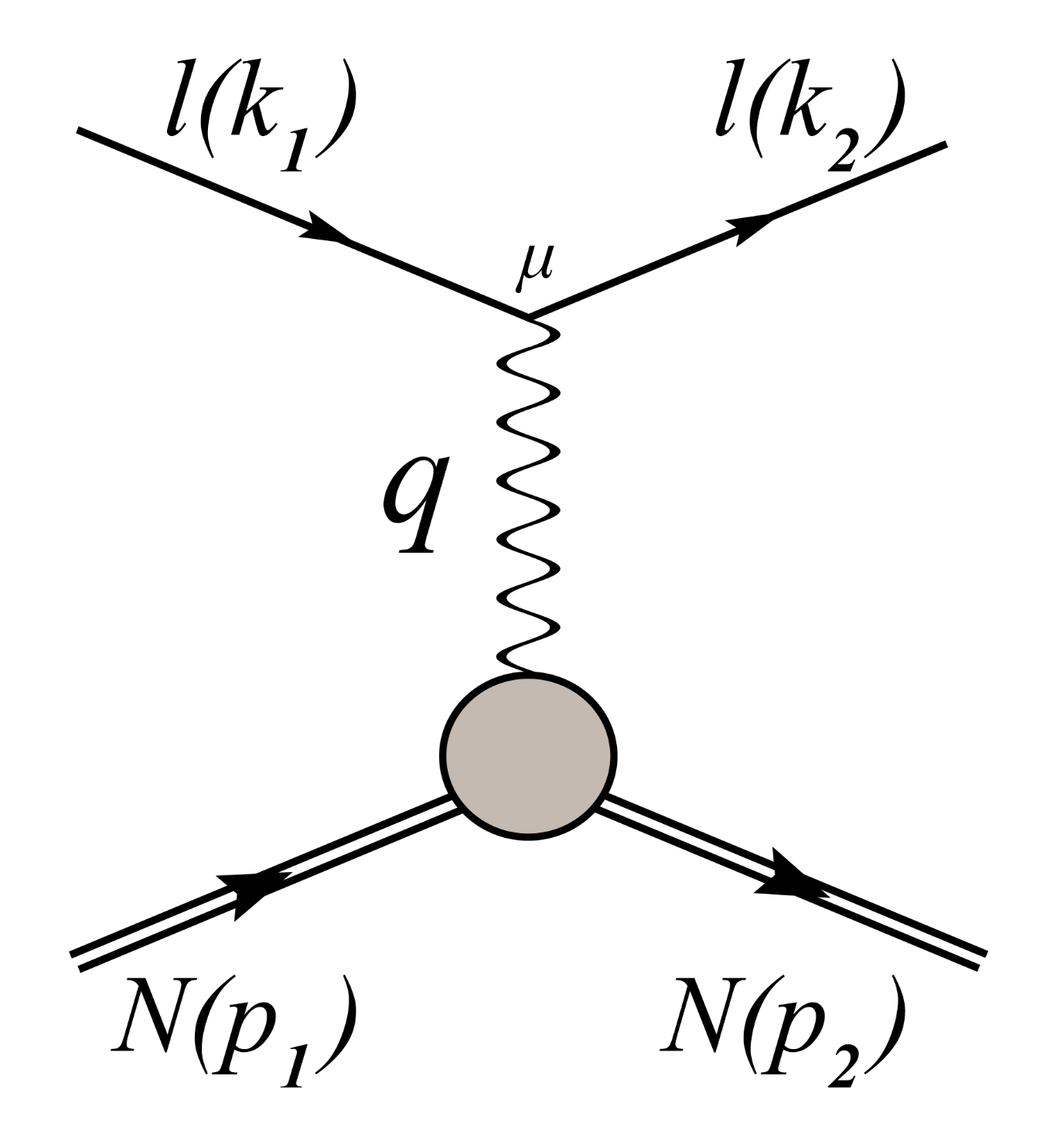}
        }\\[2mm]
        (a)
    \end{minipage}
    \hfill
    \begin{minipage}[t]{0.54\columnwidth}
        \centering
        \scalebox{1}[1.0]{%
            \includegraphics[width=\linewidth]{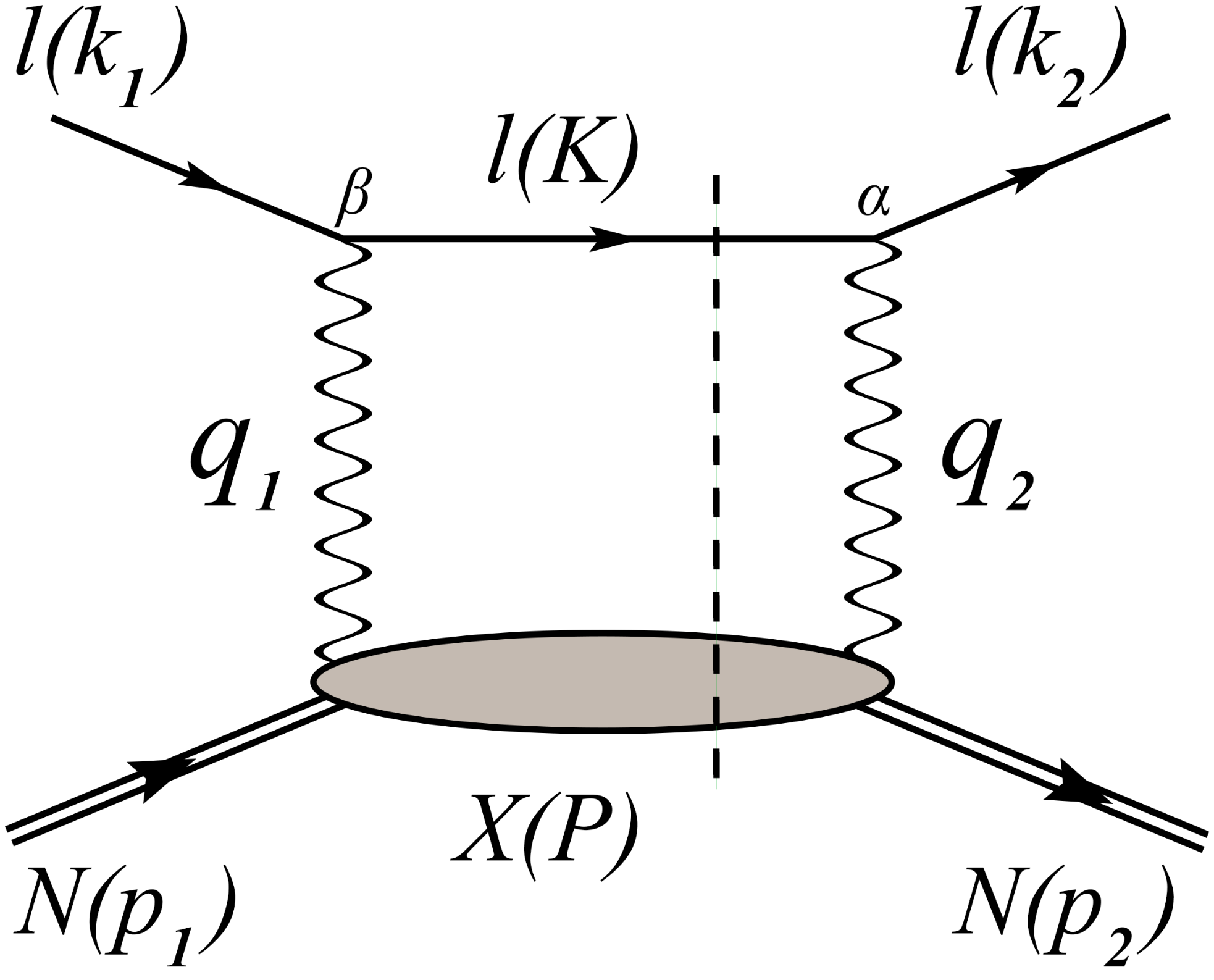}
        }\\[2mm]
        (b)
    \end{minipage}

    \caption{(a) One-photon, $M_{1\gamma}$ and (b) two-photon, $M_{2\gamma}$ exchange diagrams for elastic lepton-nucleus scattering}
    \label{fig:diagrams}
\end{figure}
%%%%%%%%%%%%%%%%%%%%%%%%%%%%%%%%%%

\section{BNSSA in elastic lepton-nucleus scattering}

The elastic scattering of leptons from spin-zero nuclei can be represented as follows
\begin{equation}
e(k_1) + N(p_1) \rightarrow e(k_2) + N(p_2),
\end{equation}
where the initial and final lepton four-momenta are given by $k_1$ and $k_2$, respectively, while $p_1$ and $p_2$ represent the corresponding initial and final four-momenta of the target nucleus.

In this scattering process, the beam-normal SSA is defined as
\begin{equation}
B_n \equiv \frac{d \sigma_{\uparrow}-d \sigma_{\downarrow}}{d \sigma_{\uparrow}+d \sigma_{\downarrow}}.
\end{equation}
Here, $d \sigma_{\uparrow}$ ($d \sigma_{\downarrow}$) is the differential cross section for scattering of leptons polarized parallel (antiparallel) to the normal vector of the scattering plane from unpolarized spin-zero nuclei. For an incoming lepton, the corresponding normal spin four-vector is given by
\begin{equation}
S_l^\mu=\frac{1}{N_s} \varepsilon_{\mu \nu \rho \sigma} p_1^\nu k_1^\rho k_2^\sigma,
\end{equation}
where the normalization factor $N_s$ is chosen such that $S_l^2=-1$. Explicitly, it is defined as
\begin{equation}
N_s=\frac{1}{2} \sqrt{-t\left[\left(M^2-s\right)^2 + s t-2 m^2\left(M^2+s\right)+m^4\right]},
\end{equation}
where $s = (k_1 + p_1)^2$, and $t = (k_1 - k_2)^2$ are the standard Mandelstam variables. Constants $m$ and $M$ denote the masses of the lepton and the target nucleus, respectively.

In order to derive a perturbative expression for the BNSSA, one can adapt the formalism from Ref.\bluecite{derujula1971} to the case of an unpolarized target and a transversely polarized beam. As a result, the leading-order BNSSA can be written as
\begin{equation}
B_n =\frac{\operatorname{Im}\left(\sum_{\text {spins }} M_{1\gamma}^* \operatorname{Abs}\left[M_{2 \gamma}\right]\right)}{\sum_{\text {spins }}\left|M_{1\gamma}\right|^2},
\label{asymmetry-definition-2}
\end{equation}
where $M_{1\gamma}$ and $M_{2\gamma}$ are the one- and two-photon-exchange amplitudes shown in Fig. \ref{fig:diagrams}. It is important to note that, at the Born level, the denominator in Eq.~\eqref{asymmetry-definition-2} is proportional to the differential cross section for unpolarized elastic lepton–nucleus scattering and can be expressed as
\begin{equation}
\sum_{\text{spins}} |M_{1\gamma}|^2 = \frac{32 [\alpha^2 Z]F(Q^2)}{  Q^4} D(s,t) ,
\end{equation}
where $Z$ is the nuclear charge, $Q^2  = -t$, $F(Q^2)$ is the nuclear charge form factor, and the kinematic factor $D(s,t)$ is defined as 
\begin{equation}
D(s,t) = 2\pi^2 ZF(Q^2) ((m^2+M^2-s)^2+(s-m^2) t).
\end{equation}

After  calculating the absorptive part of $M_{2\gamma}$ using the Cutkosky cutting rules, a compact expression for $B_n$ (in the CM frame) in terms of the leptonic $L_{\mu\alpha\beta}$ and hadronic $H^{\mu\alpha\beta}$ tensors can be derived
\begin{equation}
B_n = \frac{\alpha t}{D(s,t)}\int{}{} \frac{d^3\mathbf{K}}{2E_K} \frac{L_{\mu\alpha\beta}H^{\mu\alpha\beta}}{q_1^2q_2^2}, 
\end{equation}
where $\mathbf{K}$ and $E_K$ are, respectively, the three-momentum and energy of the intermediate lepton in the CM frame. In the expression above, $L_{\mu\alpha\beta}$ and $H^{\mu\alpha\beta}$ are given by
\begin{equation}
L_{\mu\alpha\beta} = \frac{1}{2}\operatorname{Tr}[(\slashed{k}_1 + m)(1 - \gamma_5\slashed{S}_{l})\gamma_{\mu}(\slashed{k}_2 + m)\gamma_{\alpha}(\slashed{K} + m)\gamma_{\beta}],
\label{BNSSA-3}
\end{equation}
\begin{equation}
H^{\mu\alpha\beta} = \frac{1}{2}\operatorname{Im}(T^{\beta \alpha})  (p_1 + p_2)^{\mu},
\label{hadronic-tensor-1}
\end{equation}
where $T^{\beta \alpha}$ is a non-forward Compton tensor for the spin-zero target.

For a spin-zero target, the most general form of $T^{\beta \alpha}$ contains five independent invariant structures derived in \bluecite{tarrach1975}. The Compton amplitude is determined using the optical theorem, which connects its value in the forward limit to the total photoabsorption cross section $\sigma_{\gamma N}$. As a result, we keep only those terms in $T^{\beta \alpha}$ that survive in the limit $q^2\rightarrow0$ and $q_1^2 \rightarrow 0$ ( which automatically implies that $q_2^2 \rightarrow 0$). Under these constraints, the Compton tensor takes the form
\begin{equation}
\begin{aligned}
& T^{\beta \alpha}=\left[-(\bar{p} \bar{q})^2 g^{\alpha \beta}-\left(q_1 q_2\right) \bar{p}^\alpha \bar{p}^\beta+(\bar{p} \bar{q})\left(\bar{p}^\beta q_{1}^{ \alpha}+\right.\right. \\
& \left.\left.\bar{p}^\alpha q_{2}^ {\beta}\right)\right] A\left(q_1^2, q_2^2, q^2, W^2\right), \\
& \bar{p}=\frac{1}{2}\left(p_1+p_2\right), \quad \bar{q}=\frac{1}{2}\left(q_1+q_2\right),
\label{eq-Compton-Tensor}
\end{aligned}
\end{equation}
where $W$ is the invariant mass of the intermediate hadronic system.

The scalar function $A\left(q^2 = 0, q_1^2 = q_2^2, W^2\right)$ in Eq.~\eqref{eq-Compton-Tensor} is normalized such that its imaginary part is related to the structure function $W_1(W^2,q_1^2)$ through
\begin{equation}
\begin{aligned}
&\operatorname{Im} \left[{A}(q^2 = 0, q_1^2 = q_2^2, W^2)\right] =\\
&= \frac{8 \pi}{(W^2 - M^2 - q_1 q_2)^2}   W_1(W^2, q_1^2),
\end{aligned}
\label{ImA}
\end{equation}
where $W_1$ can be expressed in terms of the total photoabsorption cross section as
\begin{equation}
W_1\left(W^2, 0\right)=\frac{W^2-M^2}{8 \pi^2 \alpha} \sigma^{\mathrm{tot}}_{\gamma A}\left(W^2\right).
\end{equation}

The dominant contribution to the beam-normal SSA comes from the phase-space region where the virtualities of both exchanged photons are small. Therefore, using Eqs.~\eqref{hadronic-tensor-1}, ~\eqref{eq-Compton-Tensor}, and ~\eqref{ImA}, one can show that $H^{\mu\alpha\beta}$ can be written as
\begin{equation}
H^{\mu \alpha \beta}=2 \pi W_1 (p_1 + p_2)^{\mu}\left(-g^{\alpha \beta}-\frac{2\left[p_1 q\right]^{\alpha \beta}}{W^2-M^2 - q_1q_2}\right),
\end{equation}
where we have ignored the terms proportional to $q_1q_2$, since their contribution to the BNSSA is of order $Q^2/W^2$. 

It is important to note that the equations above allow us to calculate the BNSSA only at zero scattering angle. The extrapolation of the results to nonzero angles is discussed in the next section.
%%%%%%%%%%%%%%%%%%%%%%%%%%%%%%%%%%

\section{Compton amplitude in the nonforward limit}

The optical theorem allows us to determine the imaginary part of the two-photon-exchange amplitude only in the forward limit. Ideally, its extension to nonzero scattering angles would be constrained by the slope of the differential Compton cross section. Unfortunately, the available experimental data remain very limited, so additional approximations are required. One previous approach \bluecite{gorchtein2008} replaced the ratio of the nuclear Compton form factor to the nuclear charge form factor by the analogous ratio for a proton target. We will show that for actual electric form factors and realistic models of Compton amplitudes, high-Z nuclei (like $^{208}$Pb) develop oscillatory dependence on the transferred momenta in the kinematics of Jefferson Lab's PREX and CREX experiments. 
%Although this model gives reasonably accurate results for light nuclei, it fails to explain the near-zero positive values of the BNSSA measured for ${}^{208}_{82}\mathrm{Pb}$.

In this work, we start with an observation that in the forward limit and above the nucleon-resonance region, the total photo-absorption cross section for nucleons and nuclei is nearly constant. This phenomenon is attributed to either soft or hard diffraction - depending on energy scales - described by Pomeron exchange due to strongly-interacting hadronic content of a photon. We use vector meson dominance (VMD) to model the $t$-dependence of the hadronic part of the two-photon-exchange amplitude in terms of vector-meson--nucleus scattering. 
%At the beam energies of interest, the interaction is expected to be represented by diffractive, Pomeron-like exchange. 
In the VMD picture, the dominant vector-meson contribution is associated with the neutral rho meson, $\rho^0$. Its scattering amplitude can be estimated using the Additive Quark Model (AQM), which relates the $\rho^0$ amplitude to the corresponding $\pi^+$ and $\pi^-$ amplitudes as
\begin{equation}
F_{\rho^0} = \frac{1}{\sqrt{2}}\left(F_{\pi^+} + F_{\pi^-}\right).
\label{eq:amplitude}
\end{equation}

%This approach is motivated by the fact that, at the relevant beam energies, the process lies above the resonance region, where the total photoabsorption cross section varies slowly with energy, which is characteristic of diffractive Pomeron exchange. 
We use experimental constraints for the total photo-absorption cross sections on nuclei, while applying a diffractive model to predict Compton form factor at finite momenta. Therefore, the model should capture the main behavior of the imaginary part of the two-photon-exchange (TPE) amplitude as a function of transferred momenta.

To calculate the amplitudes $F_{\pi^+}$ and $F_{\pi^-}$, we use the eikonal model developed in Ref. \bluecite{chen1993}. To make the subsequent calculation self-contained and to establish the notation, we present the relevant elements of the eikonal formulation below.

In the center-of-mass frame, the differential cross section is given by the squared magnitude of the scattering amplitude $F(q)$:
\begin{equation}
\frac{d\sigma}{d\Omega} = |F(q)|^2 ,
\end{equation}
where $q$ is the momentum transfer: $q = 2k_\pi \sin(\theta/2)$; $\theta$ is the CM scattering angle and $k_\pi$ is the pion momentum in the pion-nucleus CM frame. 

The total scattering amplitude consists of a point-charge Coulomb contribution, $F_{\mathrm{pt}}(q)$, and a Coulomb-modified nuclear amplitude, $F_{C N}(q)$
\begin{equation}
F(q)=F_{\mathrm{pt}}(q)+F_{C N}(q).
\end{equation}
The point-charge Coulomb amplitude is known analytically,
\begin{equation}
F_{\text{pt}}(q) = -\frac{2\eta k_\pi}{q^2} \exp\left[ -i\eta \ln\left(\frac{q^2}{4k_\pi^2}\right) + 2i\sigma_0 \right], 
\end{equation}
where $\sigma_0 = \arg \Gamma(1 + i\eta)$ represents the the s-wave Coulomb phase shift. The interaction strength is governed by the Sommerfeld parameter, $\eta  = Z_{\pi} Z_{A} \alpha E/k_\pi$, with 
 $Z_\pi = \pm 1$ denoting the pion charge, $Z_A$ the target proton number, and $E$ the total center-of-mass energy.

The nuclear contribution is evaluated within the eikonal approximation via an integration over the impact parameter $b$,
\begin{equation}
F_{C N}(q)=i k_\pi \int_0^{\infty} b d b J_0(q b) e^{i \chi_{\mathrm{pt}}(b)} \Gamma_{C N}(E, b).
\end{equation}
In this integral, the rapid oscillations of the point Coulomb phase,  $\chi_{\text{pt}}(b) = 2\eta \ln(k_\pi b)$, are explicitly factored out. The physics of the collision is therefore contained within the profile function, $\Gamma_{C N}$, which describes how the incident wave is attenuated and phase-shifted as it passes through the nucleus
\begin{equation}
\begin{aligned}
\Gamma_{C N}(E, b) =  & 1-\exp \left(i \chi_N\left[E, b\left(1+E V_C(b) / k_\pi^2\right)\right]\right. \\
& \left.+i \chi_C(b)-i \chi_{\mathrm{pt}}(b)\right).
\end{aligned}
\end{equation}
The expression above separates the nuclear phase shift, $\chi_N$, from the finite-size Coulomb phase correction, $\left[\chi_C(b)-\chi_{\mathrm{pt}}(b)\right]$. $V_C(r)$ is the finite-charge Coulomb Potential, which models the target nucleus as a uniformly charged sphere of radius $R_C$.

The phase shifts are calculated by integrating their respective potentials along the longitudinal coordinate $z$. The finite-size Coulomb correction is given by
\begin{equation}
\begin{aligned}
\chi_C(b)-\chi_{\text{pt}}(b)
&= -\frac{E}{k_\pi}\int_{-\infty}^{\infty}
\left[
V_C\left(\sqrt{b^2+z^2}\right)\right. \\
&\qquad\left.
-\frac{Z_\pi Z_A\alpha}{\sqrt{b^2+z^2}}
\right] dz ,
\end{aligned}
\end{equation}
while the strong nuclear phase shift is generated by an effective optical potential $U_{\text {eik }}$
\begin{equation}
\chi_N(E, b)=-\frac{1}{2 k_\pi} \int_{-\infty}^{\infty} d z U_{\text {eik }}\left[E-V_C(r), r\right].
\end{equation}

In order to achieve high precision, the strong optical potential also includes the first Wallace correction, which takes into account the deviations from straight-line trajectories and energy variations along the path:
\begin{equation}
U_{\mathrm{eik}}(E, r)=U_0+\frac{U_0^2}{4 k_\pi^2}\left(1+\frac{2 b^2}{r} \frac{d}{d r} \ln U_0\right),
\end{equation}
where $U_0$ is a local potential constructed using pion-proton, $f_p(0)$, and pion-neutron, $f_n(0)$, scattering amplitudes as well as their derivatives in the forward limit in the pion-nucleus CM frame. Analytically, it is defined as
\begin{equation}
\begin{aligned}
U_0(E, r)= & -4 \pi Z\left[f_p(0) \rho_p+\frac{f_p^{\prime}(0)}{2 k_\pi^2} \nabla^2 \rho_p\right] \\
& -4 \pi N\left[f_n(0) \rho_n+\frac{f_n^{\prime}(0)}{2 k_\pi^2} \nabla^2 \rho_n\right],
\end{aligned}
\end{equation}
where $Z$ is the proton number, $N$ is the number of neutrons; $\rho_p$ and $\rho_n$ are the density distributions of protons and neutrons, respectively. 

The above expressions show that the final result depends on the chosen proton and neutron density distributions. In our calculations, we tested four charge-density parameterizations, along with their corresponding derivatives and form factors: Woods-Saxon, three-parameter Fermi, Fourier-Bessel, and Sum of Gaussians. The four parametrizations lead to very similar results, with differences below about $5 \%$. For this reason, the results presented below are presented using the Fourier-Bessel parametrization.

The final ingredients required to construct the optical potential are the elementary pion-nucleon scattering amplitudes and their derivatives. Since the strong  interaction conserves isospin, the physical $\pi p$ and $\pi n$ scattering amplitudes can be directly constructed from the isospin $I = \frac{3}{2}$ and $I = \frac{1}{2}$ states:
\begin{equation}
f(\pi^+p) = f^{3/2},
\end{equation}
\begin{equation}
f(\pi^-p) = \bra{\pi^-p}\hat{F}\ket{\pi^-p} = \frac{1}{3}f^{3/2} + \frac{2}{3}f^{1/2}.
\end{equation}
$f^{3/2}$ and $f^{1/2}$ are the scattering amplitudes for the states with total isospin $I=3/2$ and $I=1/2$, respectively. 

Since the strong interaction is invariant under the exchange of $u$ and $d$ quarks as well as $\bar{u}$ with $\bar{d}$ antiquarks (Charge Symmetry), one can easily establish the following relations for the pion-neutron scattering amplitudes
\begin{equation}
\begin{aligned}
f(\pi^+n) &= f(\pi^-p), \\
f(\pi^-n) &= f(\pi^+p).
\end{aligned}
\end{equation}

In order to construct the local potential $U_0$, we need to know these amplitudes and their angular derivatives calculated at $\theta = 0^{\circ}$. The forward isospin scattering amplitudes $f^I$ and their derivatives $\frac{d}{cos \theta} f^I$ can be expressed in terms of the complex partial wave amplitudes $T_{L \pm}^I$:
\begin{equation}
f^I(0)=\frac{1}{\kappa_{c m}} \sum_{L=0}^{L_{\max }}\left[(L+1) T_{L+}^I+L T_{L-}^I\right],
\end{equation}
\begin{equation}
f^{\prime I}(0)=\frac{1}{\kappa_{c m}} \sum_{L=0}^{L_{\max }}\left[(L+1) T_{L+}^I+L T_{L-}^I\right] \frac{L(L+1)}{2},
\end{equation}
where $T_{L \pm}^I$ are the complex partial wave amplitudes for a given Isospin $I$, orbital angular momentum $L$ and Total Spin $J=L \pm 1 / 2$. In this study, $T_{L \pm}^I$ were extracted from the SAID database.

The amplitudes given above are defined in the pion-nucleon ($\pi N$) CM frame. However, the optical potential operates in the pion-nucleus ($\pi A$) CM frame. As a result, the above amplitudes and their derivatives must be kinematically transformed to the ($\pi A$) CM frame. For a given target nucleon $N$ (where $N$ corresponds to either a proton $p$ or a neutron $n$ ), the forward scattering amplitude and its angular derivative transform as:
$$
\begin{aligned}
f_N(0) & =\frac{k_\pi}{\kappa} f(\pi N)|_{\theta = 0}, \\
f_N^{\prime}(0) & =\left(\frac{k_\pi}{\kappa}\right)^3 f^{\prime}(\pi N)|_{\theta = 0},
\end{aligned}
$$
where $f(\pi N)|_{\theta = 0}$ is the corresponding forward scattering amplitude in the pion-nucleon CM frame; $k_{\pi}$ and $\kappa$ are the magnitudes of the pion 3-momenta in the pion-nucleus CM frame and pion-nucleon CM frame, respectively.

The discussed approach was tested against the experimental data for $\pi^+$ and $\pi^-$ differential scattering cross-sections on ${}^{12}_{6}\mathrm{C}$ and ${{}^{208}_{82}}\mathrm{Pb}$. As one can see from Figs.~\ref{fig:pion-scattering-carbon} and~\ref{fig:pion-scattering-lead} the predicted cross-section values are slightly smaller; however, the model does manage to accurately describe the trend of the curves and precisely determines the position of diffraction minima. 

For our purposes, we are interested in the imaginary part of the doubly virtual Compton amplitude, which we assume should exhibit the same behaviour as the $\rho^0$ amplitude at nonzero $t$. The latter is determined from the Eq.~\eqref{eq:amplitude}, where the imaginary parts of $F_{\pi^+}$ and $F_{\pi^-}$ were calculated when the electromagnetic interaction is turned off (since the charge of $\rho^0$ is equal to zero). To determine the Compton amplitude at nonzero $t$, we fitted the resulting $F_{\rho^0}(t)$ using the following ansatz:
\begin{equation}
\operatorname{Im}(F_{\rho^0}(t)) = A_0 \frac{2 J_1(q R)}{q R} \exp \left(-B q^2\right),
\end{equation}
where $J_1(q R)$ is the Bessel function of the first kind of order 1. $A_0$, $R$ and $B$ are the parameters determined from a fit.
As a result, our ansatz for the imaginary part of the doubly-virtual Compton tensor, $\operatorname{Im}(T^{\beta\alpha})$ must be proportional to
\begin{equation}
\operatorname{Im}(T^{\beta\alpha}) \propto \sigma^{\mathrm{tot}}_{\gamma A} (W^2, t = 0) \frac{2 J_1(q R)}{q R} \exp \left(-B q^2\right),
\end{equation}
where $\sigma^{tot}_{\gamma A} (W^2, t = 0)$ is the total nuclear photoabsorption cross-section in the forward limit, $R$ and $B$ are the parameter values determined from the previous step.

%%%%%%%%%%%%%%%%%%%%%%%%%%%%%%%%%%
\section{Results and discussion}

As mentioned above, an accurate calculation of the BNSSA requires reliable knowledge of the differential Compton-scattering cross section. At the time of writing, such information is not available over the relevant energy range. As a result, one of the most recent attempts to explain the PREX puzzle took one step further in modeling the imaginary part of the Compton amplitude \bluecite{koshchii2021}. In addition to the exponential fall-off with momentum transfer, the Compton amplitude was assumed to be proportional to the charge form factor of the nuclear target. In the context of the Born approximation, such an ansatz effectively removes the effects of diffractive minima on the inelastic part of the asymmetry
\begin{equation}
\frac{g_N(Q^2)}{F_N(Q^2)} = \frac{F_N(Q^2) \times[\text{Other nonsingular terms}]}{F_N(Q^2)}.
\end{equation}
Since the phenomenological Compton form factor was assumed in Ref.\cite{koshchii2021} to vanish at exactly the same values of the momentum transfer as the nuclear charge form factor, the resulting asymmetry has a smooth functional behavior. We believe this is a reasonable approximation for nuclei such as ${}^{12}_6\mathrm{C}$ and ${}^{40}_{20}\mathrm{Ca}$, in which the proton and neutron distributions are approximately the same. The ${}^{208}_{82}\mathrm{Pb}$ target, on the other hand, possesses a large neutron skin ($\approx 0.28  \text{ fm}$) \bluecite{adhikari2021prex} and, since both protons and neutrons respond to the two-photon probe, it suggests that the positions of diffractive minima of the Compton form factor and the nuclear charge form factor would not coincide in general. 

We did not make such a restrictive assumption in our approach. Instead, the positions of the diffractive minima in the Compton amplitude are determined from the calculation of the elastic $\rho^0$-nucleus scattering, with the result related to the Compton amplitude through the VMD model. The findings presented below suggest that it is the mismatch in the positions between the minima of the Compton amplitude and those of the nuclear charge form factor that might explain the different sign and the measured low values of the asymmetry for $^208$Pb target.

%%%%%%%%%%%%%%%%%%%%%%%%%%%%%%%%%%
\begin{figure}[htbp]
    \capstart
    \centering
    \includegraphics[width=\columnwidth]{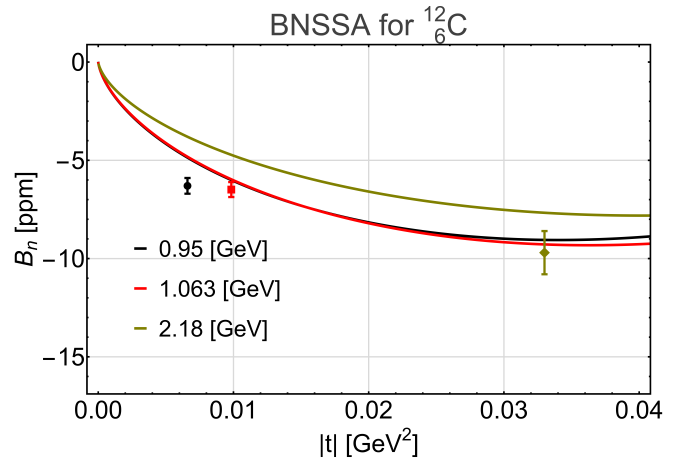}
    \caption{Beam-normal single-spin asymmetry $B_n$ (in parts per million) for ${}^{12}_{6}\mathrm{C}$ as a function of $|t|$, calculated at beam energies of $0.95$, $1.063$, and $2.18~\mathrm{GeV}$. The corresponding fitted parameter values $(B,R)$ are $(13.0~\mathrm{GeV}^{-2}, 2.6~\mathrm{fm})$, $(13.2~\mathrm{GeV}^{-2}, 2.6~\mathrm{fm})$, and $(13.3~\mathrm{GeV}^{-2}, 2.5~\mathrm{fm})$, respectively.
 The corresponding experimental data points are taken from Refs.~\bluecite{abrahamyan2012bnssa,adhikari2022}.}
    \label{fig:asymmetry carbon}
\end{figure}

%%%%%%%%%%%%%%%%%%%%%%%%%%%%%%%%%%

The results for the BNSSA ($B_n$) for the $^{12}_6\text{C}$ as a function of the squared momentum transfer ($|t|$) are presented in Fig.~\ref{fig:asymmetry carbon}. The theoretical curves were obtained at beam energies corresponding to the measurements conducted by HAPPEX and PREX Collaborations and by the PREX and CREX Collaborations: $0.95$, $1.063$, and $2.18 \text{ GeV}$. Since $^{12}_{6}\text{C}$ is a light, symmetric nucleus ($N=Z$), its proton and neutron density distributions are virtually identical. As a result, $F_N(Q^2)$ and $g_N(Q^2)$ remain highly correlated in the kinematic regimes under study, which results in the smooth, monotonically increasing negative asymmetries shown in the figure. Our VMD-based approach successfully reproduces the sign and the overall trend of the asymmetry, as well as the experimental result at 1.063 GeV. It is important to point out, however, that the model somewhat underestimates the result at the other two energies. These discrepancies are most likely caused by a different fall-off behavior of the Compton amplitude in the resonance region compared to the one at higher energies.

%%%%%%%%%%%%%%%%%%%%%%%%%%%%%%%%%%
\begin{figure}[htbp]
    \capstart
    \centering
    \includegraphics[width=\columnwidth]{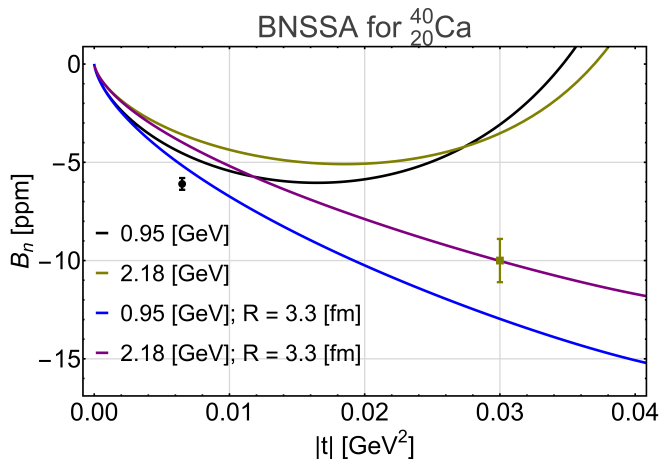}
    \caption{Beam-normal single-spin asymmetry $B_n$ (in parts per million) for ${}^{40}_{20}\mathrm{Ca}$ as a function of $|t|$, calculated at beam energies of $0.95$ and $2.18~\mathrm{GeV}$. The fitted parameter pairs $(B,R)$ are $(15.5~\mathrm{GeV}^{-2}, 4.0~\mathrm{fm})$ at $0.95~\mathrm{GeV}$ and $(15.8~\mathrm{GeV}^{-2}, 3.9~\mathrm{fm})$ at $2.18~\mathrm{GeV}$. The corresponding experimental data points are taken from Ref.~\bluecite{adhikari2022}. The two additional curves—the blue curve at $0.95~\mathrm{GeV}$ and the purple curve at $2.18~\mathrm{GeV}$—correspond to calculations with $R=3.3~\mathrm{fm}$. }
    \label{fig:asymmetry calcium}
\end{figure}

%%%%%%%%%%%%%%%%%%%%%%%%%%%%%%%%%%

Fig.~\ref{fig:asymmetry calcium} displays the BNSSA results for ${}^{40}_{20} \mathrm{Ca}$, another doubly magic nucleus. The asymmetry was calculated at beam energies of 0.95 and 2.18 GeV. As shown in the figure, there is a noticeable mismatch between the theoretical results and the experimental data. The theoretical curves exhibit an early turnaround and consequently systematically underestimate the measured asymmetry at the corresponding values of the momentum transfer. The disagreement most likely stems from the values of the effective hadronic radius, $R$, determined by the model, which sets it equal to approximately $4$ fm. Since the estimated value is larger than the corresponding value of the charge radius, $R_{\mathrm{ch}}$ of  ${}^{40}_{20} \mathrm{Ca}$, the zero of the Compton amplitude gets shifted with respect to the zero of the charge form factor. This mismatch of nuclear response to single-photon and two-photon probe prevents the zero-crossings from canceling out in the $g_N(Q^2)/F_N(Q^2)$ ratio, which results in a sharp curvature in the predicted asymmetry. 

It was established that,  ${}^{40}_{20} \mathrm{Ca}$ nucleus has virtually no neutron skin similar to $^{12}_6\text{C}$, \bluecite{zenihiro2018ca}, which implies that its proton and neutron distributions are approximately the same. It suggests that the effective radius entering the Compton amplitude should have approximately the same value as $R_{\mathrm{ch}}$ for these lighter nuclei. In order to test the theory under this assumption, we calculated two additional curves at 0.95 and 2.18 GeV (represented by the olive and purple curves, respectively) for which the value of $R$ was set to $3.3$ fm. This modification makes the theoretical results flatten out and increase monotonically, similar to the case of the $^{12}_6\text{C}$ target. The results, particularly the one at $2.18$ GeV, also show better agreement with the experimental data.

Evidently, our predictions are highly sensitive to the extracted values of $R$. The deviation of the modeled effective radius from the physical radius reflects a known limitation of applying VMD and Pomeron-exchange models, which assume high-energy coherent diffractive scattering at intermediate energies. At the beam energies under consideration, the Compton amplitude also contains contributions from incoherent quasi-elastic scattering and from the excitation of individual nucleon resonances (such as the $\Delta(1232)$). Unfortunately, the uncertainty introduced to the theory by these effects is challenging to estimate, and a detailed discussion of their contributions should be addressed in future study.

%%%%%%%%%%%%%%%%%%%%%%%%%%%%%%%%%%

\begin{figure}[t!]
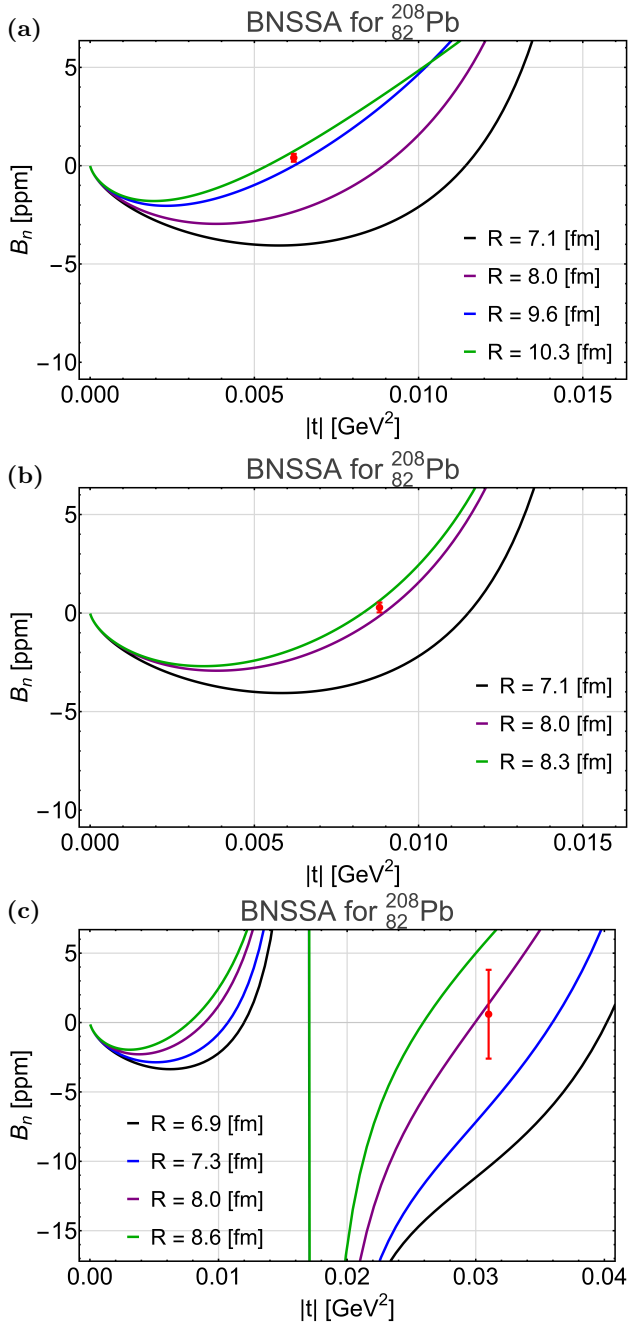

    \capstart
    \centering

    \begin{overpic}[width=0.95\columnwidth]{FigBnTPb208_Our_fit_095.png}
        \put(0,65){\textbf{(a)}}
    \end{overpic}

    \vspace{0.4em}

    \begin{overpic}[width=0.95\columnwidth]{FigBnTPb208_Our_fit_1063.png}
        \put(0, 65){\textbf{(b)}}
    \end{overpic}

    \vspace{0.4em}

    \begin{overpic}[width=0.95\columnwidth]{FigBnTPb208_Our_fit_218.png}
        \put(0, 65){\textbf{(c)}}
    \end{overpic}

    \caption{%
       BNSSA for ${}^{208}_{82}\mathrm{Pb}$ as a function of $|t|$, showing its dependence on the radius parameter $R$. Panels (a), (b), and (c) correspond to beam energies of $0.95$, $1.063$, and $2.18~\mathrm{GeV}$, respectively. The fitted values of $B$ used in panels (a), (b), and (c) are $15.6$, $15.0$, and $16.0~\mathrm{GeV}^{-2}$, respectively.
 The black curves represent the results predicted by the model and use the reference values $R=7.1$, $7.1$, and $6.9~\mathrm{fm}$ in panels (a), (b), and (c), respectively. The alternative values are: (a) $R=8.0$, $9.6$, and $10.3~\mathrm{fm}$ for the purple, blue, and green curves; (b) $R=8.0$ and $8.3~\mathrm{fm}$ for the purple and green curves; and (c) $R=7.3$, $8.0$, and $8.6~\mathrm{fm}$ for the blue, purple, and green curves. The data points were taken from Refs. ~\bluecite{abrahamyan2012bnssa,adhikari2022}.
    }
     \label{fig:asymmetry lead}
    
\end{figure}

%%%%%%%%%%%%%%%%%%%%%%%%%%%%%%%%%%

\begin{figure}[!t]
    \capstart
    \centering
    \includegraphics[width= 0.95\columnwidth]{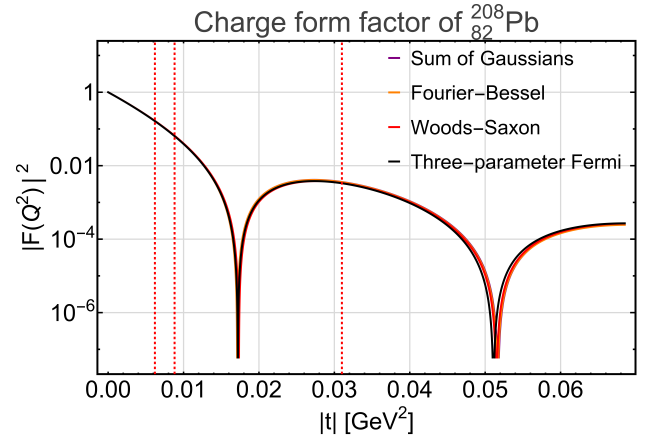}
    \caption{A model \bluecite{duda2007} charge form factor squared of $^{208}_{82}\mathrm{Pb}$ as a function of the four-momentum transfer $|t|$ used in the present study. The results are obtained using four parametrizations of the nuclear charge distribution: sum of Gaussians, Fourier--Bessel, Woods--Saxon, and three-parameter Fermi. The vertical red dashed lines indicate the $|t|$ values at which the BNSSA was measured.}
    \label{fig:form factor}
\end{figure}

Finally, we discuss the results for the ${}^{208}_{82}\mathrm{Pb}$ target. Fig.~\ref{fig:asymmetry lead}  shows our calculations of the BNSSA for the beam energies used in experiments. At all three energies, our model predicts the first and second diffractive minima in the Compton amplitude at $|t|\approx 0.012 \text{ GeV}^2$ and $|t|\approx 0.04 \text{ GeV}^2$, respectively. These minima together with the first diffractive minimum in the charge form factor (see Fig.~\ref{fig:form factor}) suggest that the leading term in the asymmetry can change sign at least twice, and potentially three times, within the experimentally studied region ($|t| \in [0  \text{ GeV}^2,0.031 \text{ GeV}^2]$ for ${}^{208}_{82}\mathrm{Pb}$). Consequently, if $R$ is treated as a free parameter in our model, this oscillatory behavior can potentially explain most of the experimental observations. For example, when $R$ is set to a value slightly larger than the one predicted by the model, namely ($R = 8 \text{ fm}$), the curves reproduce the results at 1.063 GeV and 2.18 GeV (see Fig.~\ref{fig:asymmetry lead} (b) and (c)). Admittedly, the asymmetry still does not vanish at 0.95 GeV, and $R$ has to be set to $\approx 10 \text{ fm}$ for that to be the case. However, for $R = 8 \text{ fm}$ the asymmetry is $\approx -2.3 \text{ ppm}$ at the experimental $|t|$ value, which is substantially smaller in magnitude than the values predicted by previous studies \bluecite{gorchtein2008, koshchii2021}. At this point, it is important to note that the vertical line at $|t|\approx 0.017 \text{ GeV}^2$ corresponds to the point where the charge form factor is zero. Since the asymmetry in our calculation was determined to the lowest non-vanishing order, the divergence at that point is unphysical. Higher-order effects are expected to smooth out the curve. Together with the uncertainties in the Compton amplitude in the resonance region, these effects may explain why the value of $R$ required to reproduce the experimental results should be slightly larger than that predicted in our VMD approach. A detailed study of these corrections is postponed to the next paper.

%%%%%%%%%%%%%%%%%%%%%%%%%%%%%%%%%%
\section{Conclusions}

In this study, we proposed a diffractive approach for calculating the momentum-transfer dependence of the imaginary part of the Compton amplitude to evaluate BNSSA in elastic electron-nucleus scattering. At the beam energies discussed in this study, the total photoabsorption cross section varies slowly with energy, suggesting that the Compton amplitude should exhibit diffractive behavior, which can be captured by considering hadronic ($\rho^0$) content of exchanged photons. The model predicts that the zeros in the nuclear Compton form factor for heavy nuclei, such as $^{208}_{82}\mathrm{Pb}$, do not coincide with the zeros of the corresponding charge form factor. Since BNSSA is proportional to the ratio of Compton-to-charge form factors, this misalignment produces oscillatory behavior of the BNSSA over the considered momentum-transfer range and may explain the small positive asymmetry values measured at JLab \bluecite{abrahamyan2012bnssa, adhikari2022}. 

In view of the predicted oscillatory behavior of BNSSA for heavy nuclei, and taking into account energy dependence for positions of diffractive minima, experimental verification of this possibility would need measurements of BNSSA for a given nucleus at a fixed energy and several scattering angles. Auxiliary measurements might include both virtual and real Compton scattering on nuclei in the kinematics of interest in order to directly measure the Compton form factors.

Finally, we should emphasize that the presented method applies to low-angle electron scattering with energies above the nuclear resonance region that might contribute to the two-photon loop integral. Its applicability is questionable for the kinematics of the recent measurement at MAMI \bluecite{esser2025pb} due to potentially large hadronic resonance contributions that require a non-diffractive theoretical description similar to \cite{pasquini2004}.

%%%%%%%%%%%%%%%%%%%%%%%%%%%%%%%%%%%

\FloatBarrier

\section*{Acknowledgements}

This work was supported by National Science Foundation's IMPRESS-U Program Award OISE-2433872.

%%%%%%%%%%%%%%%%%%%%%%%%%%%%%%%%%%%

%%%%%%%%%%%%%%%%%%%%%%%%%%%%

\appendix

\section{Differential scattering cross section of pions from different nuclei}

\setcounter{figure}{0}
\renewcommand{\thefigure}{A\arabic{figure}}
\renewcommand{\theHfigure}{A.\arabic{figure}}

%%%%%%%%%%%%%%%%%%%%%%%%%%%%%%%%%%
\begin{figure}[H]
    \capstart
    \centering
    \includegraphics[width=\columnwidth]{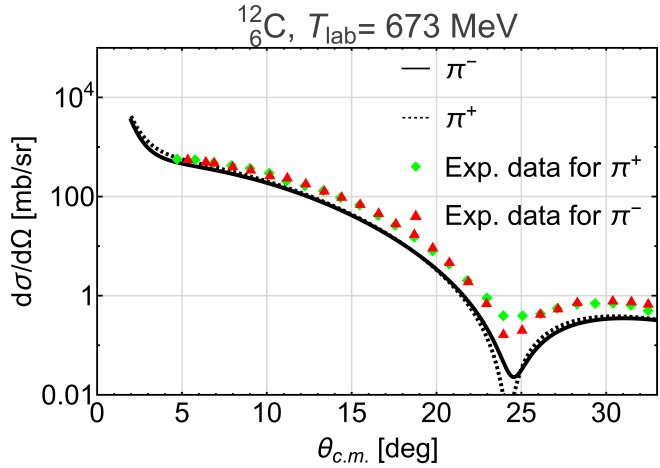}
    \caption{Center-of-mass angular distribution of the differential cross section for $\pi^\pm$ scattering from ${}^{12}_{6}\mathrm{C}$ at $T_{\mathrm{lab}}=673~\mathrm{MeV}$. The eikonal-model results for $\pi^-$ (solid line) and $\pi^+$ (dashed line). Green diamond-shaped and red triangle-shaped points represent the experimental results for $\pi^+$ and $\pi^-$, respectively. The experimental data were taken from \bluecite{marlow1984}.}
    \label{fig:pion-scattering-carbon}
\end{figure}

%%%%%%%%%%%%%%%%%%%%%%%%%%%%%%%%%%

%%%%%%%%%%%%%%%%%%%%%%%%%%%%%%%%%%
\begin{figure}[H]
    \capstart
    \centering
    \includegraphics[width=\columnwidth]{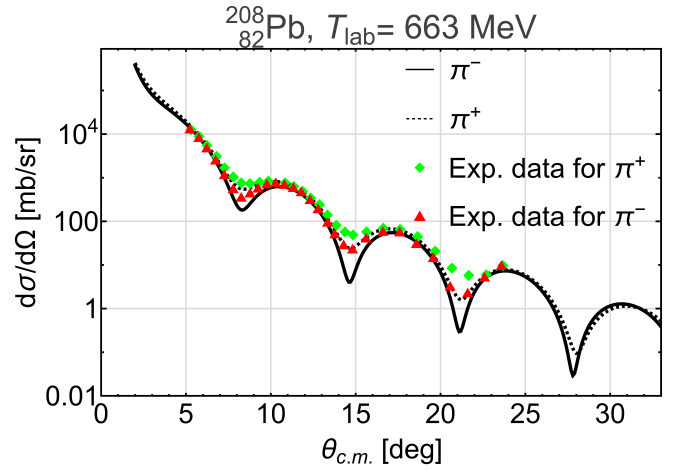}
    \caption{Center-of-mass angular distribution of the differential cross section for $\pi^\pm$ scattering from ${}^{208}_{82}\mathrm{Pb}$ at $T_{\mathrm{lab}}=663~\mathrm{MeV}$. Our numerical results for $\pi^-$ (solid line) and $\pi^+$ (dashed line) are compared with the corresponding experimental data \bluecite{kahrimanis1997}.}
    \label{fig:pion-scattering-lead}
\end{figure}

%%%%%%%%%%%%%%%%%%%%%%%%%%%%%%%%%%

%%%%%%%%%%%%%%%%%%%%%%%%%%%%

%%%%%%%%%%%%%%%%%%%%%%%%%%%%%%%%%%%

\end{document}